\documentclass[twoside]{article}
\usepackage{fleqn,espcrc2}
\usepackage{epsfig} 
\input{psfig}

\usepackage{graphicx}
\usepackage[figuresright]{rotating}

\def \ms {{\overline{\mbox{MS}}}}
\newcommand{\z}{&&\hspace*{-0.7cm}}
\newcommand{\zz}{&&\hspace*{-0.3cm}}
\newcommand{\AmS}{{\protect\the\textfont2
  A\kern-.1667em\lower.5ex\hbox{M}\kern-.125emS}}

\title{
$Q^2$ evolution of parton distributions at small $x$}

\author{A.V. Kotikov\address[MCSD]{
Particle Physics Laboratory,
Joint Institute for Nuclear Research,~
141980 Dubna, Russia}\thanks{kotikov@sunse.jinr.ru}
        and
        G. Parente\address{Dep.
de F\'\i sica de Part\'\i culas, Univ.
de Santiago de Compostela,
15706 Santiago de Compostela, Spain}\thanks{gonzalo@fpaxp1.usc.es}}
\begin{document}

\begin{abstract}
We investigate  the  $Q^2$ evolution of parton distributions
at small $x$ values,
%present an analytical parametrization of the QCD
%description of the  behaviour
%of parton distribution functions in the leading twist approximation
%of the Wilson operator product expansion,
%recently 
obtained in the case
of flat initial conditions.
The results are in excellent agreement with deep inelastic scattering
experimental data from HERA.
\vspace{1pc}
\end{abstract}

\maketitle

\section{Introduction}

The measurements of the deep-inelastic scattering
%(DIS)
structure function
%(SF)
$F_2$ in HERA
%\cite{H1,ZEUS,ZEUSB}
\cite{H1}
have permitted the access to
a very interesting kinematical range for testing the theoretical
ideas on the behavior of quarks and gluons carrying a very low fraction
of momentum of the proton, the so-called small $x$ region.
In this limit one expects that
non-perturbative effects may give essential contributions. However, the
reasonable agreement between HERA data and the 
%NLO 
next-to-leading order (NLO)
approximation of
perturbative
QCD that has been observed for $Q^2 > 1 $GeV$^2$ (see the recent review
in \cite{CoDeRo}) indicates that
%and, thus,
perturbative QCD could describe the
evolution of structure functions up to very low $Q^2$ values,
traditionally explained by soft processes.
It is of fundamental importance to find out the kinematical region where
the well-established perturbative QCD formalism
can be safely applied at small $x$.

The standard program to study the small $x$ behavior of
quarks and gluons
is carried out by comparison of data
with the numerical solution of the
%Dokshitzer-Gribov-Lipatov-Altarelli-Parisi (DGLAP)
DGLAP
equations 
%\cite{DGLAP1, DGLAP}
%\footnote{ At small $x$ there is another approach
%based on the
%%Balitsky-Fadin-Kuraev-Lipatov (BFKL) equation \cite{BFKL},
%BFKL equation, whose
%application is out of the scope of this work.} 
by fitting the parameters of the
$x$ profile of partons at some initial $Q_0^2$ and
the QCD energy scale $\Lambda$ (see, for example, \cite{MRS,KKPS}).
%\cite{fits}-\cite{GRV}.
However, if one is interested in analyzing exclusively the
small $x$ region ($x \leq 0.01$), 
there is the alternative of doing a simpler analysis
by using some of the existing analytical solutions of DGLAP 
in the small $x$ limit (see \cite{CoDeRo} for review).
%\cite{BF1}-\cite{Munich2}.
This was done so in Ref. \cite{BF1}-\cite{Q2evo1}
where it was pointed out that the HERA small $x$ data can be
interpreted in 
terms of the so called doubled asymptotic scaling phenomenon
related to the asymptotic 
behavior of the DGLAP evolution 
discovered  in \cite{Rujula}
many years ago. 

Here we illustrate results obtained recently in
\cite{Q2evo}.
These results are the extension to the NLO QCD approximation of previous 
leading order (LO)
%LO 
studies \cite{Rujula,BF1}.
The main ingredients are:

{\bf 1.} Both, the gluon and quark singlet densities are
presented in terms of two components ($'+'$ and $'-'$)
which are obtained from the analytical $Q^2$
dependent expressions of the corresponding ($'+'$ and $'-'$)
parton distributions moments.

{\bf 2.} The $'-'$ component is constant
at small $x$, whereas the 
$'+'$ component grows at $Q^2 \geq Q^2_0$ as 
%$$\sim \exp{\left(2\sqrt{\left[
%a_+\ln \left(
%\frac{a_s(Q^2_0)}{a_s(Q^2)} \right) -
%\left( b_+ +  a_+ \frac{\beta_1}{\beta_0} \right)
%\Bigl( a_s(Q^2_0) - a_s(Q^2) \Bigr) \right] 
%\ln \left( \frac{1}{x}  \right)} \right)},
%$$
$\sim \exp{(\sigma)}$, where
$$
\sigma = 2\sqrt{(\hat d_{+}s+\hat D_{+}p)lnx},
$$
%where 
and the LO term $\hat d_+ = -12/\beta_0$ and the NLO one 
$\hat D_{\pm}=\hat d_{\pm\pm}+\hat d_{\pm}\beta_1/\beta_0$ with
$\hat d_{\pm\pm} = 412f/(27\beta_0)$. 
Here the coupling constant
$a_s=\alpha_s/(4\pi)$, 
$s=ln(\alpha (Q^2_0)/\alpha (Q^2))$ and
$p=\alpha (Q^2_0) - \alpha (Q^2)$,
$\beta_0$ and $\beta_1$ are the first two 
coefficients of QCD 
$\beta$-function and $f$ is the number of active flavors.

\section{Basical formulae
%Approach
}

Thus, our purpose
%of this article
is to demonstrate the small $x$ asymptotic
form of parton distributions
%(PD)
in the framework of the DGLAP equation starting at some $Q^2_0$ with
the flat function:
 \begin{eqnarray}
f_a (Q^2_0) ~=~
A_a ~~~~(a=q,g), \label{1}
 \end{eqnarray}
where $f_a$ are the parton distributions multiplied by $x$
and $A_a$ are unknown parameters that have to be determined from data.
Through this work at small $x$ we neglect
the non-singlet quark component.

In \cite{Q2evo} an effective method to reproduce the $x$-dependence of
parton distributions has been developed. It is based on a separation
of the singular and regular parts of the  exact solution for the
moments of
parton distributions 
and 
on the method to replace Mellin convolution by usual products 
\cite{method}. 
The method allows in simplest way to reproduce the
%leading order (LO) 
LO
results \cite{BF1} and to construct the $x$-dependence
of parton densities at 
%next-to-leading order (NLO)
NLO \footnote{From 
now on, for a quantity $k(n)$ we use the notation $\hat k $
for the coefficient in the front of the singular part
when $n \to 1$ and $\overline k(n)$ for the corresponding regular part.}:
% It has the form:
 \begin{eqnarray}
\z f_a(x,Q^2) = f_a^+(x,Q^2) + f_a^-(x,Q^2)~~~~~~~\mbox{ and }~ 
\nonumber \\
\z f_a^-(x,Q^2)= A_a^-(Q^2,Q^2_0) exp(- d_{-}(1)s 
\nonumber \\ 
\zz -D_{-}(1)p) ~+~O(x), 
%\label{9.10} \\
\nonumber \\
\z f_g^+(x,Q^2) = A_g^+(Q^2,Q^2_0)
%&\cdot & 
I_0(\sigma) exp(- \overline d_{+}(1)s 
\nonumber \\ 
\zz - \overline D_{+}(1)p)
~+~O(\rho), 
%~~\mbox { and } 
%\nonumber \\
\label{9.11} \\
\z f_q^+(x,Q^2)= 
%A_q^-  exp(- d_{-}(1)s-D^q_{-}(1)p) 
%\label{9.2} \\&+& 
A_q^+(Q^2,Q^2_0)
\biggl[ \left(1 - \bar{d}_{\pm}^q(1) \alpha(Q^2)\right) 
\nonumber \\
\zz \times \rho I_1(\sigma)
%\nonumber \\
      + 20 \alpha(Q^2) I_0(\sigma) \biggr]
\nonumber \\
\zz \times 
%&\cdot & \cdot 
exp(- \overline d_{+}(1)s-\overline D_{+}(1)p)
~+~ O(\rho),
\nonumber \\
%\label{9.12} \\
%
\z F_2(x,Q^2)= e 
%\cdot 
\biggl(f_q(x,Q^2) + \frac{2}{3}f\alpha(Q^2) f_g(x,Q^2)
 \biggr), \nonumber 
%\label{9}
\end{eqnarray}
where $I_{\nu}(\sigma)$ are modified Bessel functions, 
which have $\nu$-independent limit $exp{(\sigma)}$
at $\sigma \to \infty$,
$e=(\sum_1^f e^2_i)/f$ is the average charge square for $f$
active quarks,
%$s$ and $p$
%are given by 
%$s=ln(\alpha (Q^2_0)/\alpha (Q^2)),
%~p=\alpha (Q^2_0) - \alpha (Q^2)$, and 
%\begin{eqnarray}
$
%\sigma = 2\sqrt{(\hat d_{+}s+\hat D_{+}p)lnx} ~~, ~~~~
\rho = \sigma /2/ln(1/x) $
%\sqrt{\frac{(\hat d_{+}s+\hat D_{+}p)}{lnx}}=
%\frac{\sigma }{2ln(1/x)}
%are NLO generalizations of the corresponding Ball-Forte variables 
and the magnitudes
\begin{eqnarray}
%\label{a1}\\
\z
A_g^+(Q^2,Q^2_0) = \Bigr[1-\frac{80}{81}f\alpha(Q^2) \Bigr]A_g 
\nonumber \\
\zz + \frac{4}{9}\Bigl[1 +3(1+\frac{1}{81}f)\alpha(Q^2_0) - 
\frac{80}{81}f  \alpha(Q^2)
\Bigr] A_q, 
\nonumber \\
\z
A_g^-(Q^2,Q^2_0) = A_g - A_g^+(Q^2,Q^2_0),  \label{a2} \\
\z 
A_q^+ = \frac{f}{9}\biggl(A_g + \frac{4}{9} A_q \biggl), ~
A_q^- = A_q - 20 \alpha(Q^2_0) A_q^+
\nonumber   \end{eqnarray}
%are magnitudes of $\pm$ components.
%Wherever in this work we use the notation $\alpha = \alpha_s/(4\pi)$.
%
%
%
%
%

%
%
%
%
The 
%nonzero components of the singular and 
regular parts of the terms $d_{\pm}$
and $D_{\pm}$ have the 
form \footnote{The nonzero components of the singular parts were given in
Introduction.}:
 \begin{eqnarray}
%\hat d_{+} &=& -\frac{12}{\beta_0}, ~~~
%
\z
\overline d_{+}(1) = 1+ \frac{4}{3\beta_0}f , ~ 
%\nonumber \\
%
d_{-}(1) = \frac{16}{27\beta_0}f , 
\nonumber \\
%
%\hat d_{++} = \frac{412}{27\beta_0}f ~~, ~~~ \nonumber \\
%
%\hat d^q_{+-} &=& -20 ~~, ~~~
%
%\overline d^g_{+-}(1) = \frac{80}{81}f~~, \nonumber \\
%
\z
\overline d_{++}(1) = \frac{8}{\beta_0}
\biggl( 36 \zeta_3 + 33 \zeta_2 - \frac{1643}{12} 
\nonumber \\
\zz
+\frac{2}{9}f 
\Bigr[ \frac{68}{9} -4 \zeta_2 - \frac{13}{243}f \Big] \biggr), \nonumber \\
%
%\overline d^q_{+-}(1) &=& 
%
%\frac{134}{3} -12 \zeta_2 - \frac{13}{81}f ~~, ~~~ \nonumber \\
%
%\overline d^g_{-+}(1) &=& -3 \Bigl( 1+ \frac{f}{81} \Bigr),~~~~ ~~~
%
%\nonumber \\
%
\z
d_{--}(1) = \frac{16}{9\beta_0}
\biggl( 2 \zeta_3 - 3 \zeta_2 + \frac{13}{4} \nonumber \\
\zz + f 
\Bigr[  4 \zeta_2 - \frac{23}{18} + \frac{13}{243}f \Big] \biggr),
%\nonumber \\
%
%d^q_{-+}(1) &=& 0~~, ~~~
%
%.
 \label{9.3}
 \end{eqnarray}
where 
%$\beta_0$ is the first coefficient of QCD $\beta$-function,
$\zeta_n$ are Euler $\zeta$-functions.
% and $f$ is the number ofactive quarks.

%Looking carefully Eqs. (\ref{9.11}), we arrive to the following 
%conclusions:
%\begin{itemize}
%\item 
%We would like to note that
%our NLO results coincide with the corresponding of Ball
%and Forte in Ref. \cite{BF2} if one neglects the ``$-$'' component,
%expands our NLO singular terms
%$(\rho)^k I_{k+1}(\sigma)$ in the vicinity of the point
%$\sigma = \sigma_{LO} $ and ignores the NLO regular terms (i.e. put  
%$exp{(-\overline D_{+}(1)p)}=exp{(-D_{-}(1)p)}=1$ and cancel the terms 
%proportional to $\alpha(Q^2)$ and 
%$\alpha(Q^2_0)$ into the normalization factors $A_g^\pm$
%and $A_q^{\pm}$). We think, however, that this expansion is not so correct
%because it generates NLO corrections of the order of the LO terms.

%\item The negative sign of the NLO correction in  $\sigma$
%(see Eq. (\ref{a1}))  
%makes excellent the agreement of our result with the parametrization of $F_2$
%obtained by De Roeck and De Wolf \cite{DRDW}.
%Their result is very similar to our LO form
%of $f_q^+$ in Eq.(\ref{8.0}) if one replaces  $s_{LO} \to
%s_{LO}^{\delta}$ in the definition (\ref{2.5}) of $s_{LO}$. The
%value $\delta = 0.708$ has been obtained in the fit to H1 and ZEUS
%data. Due to $\delta <1$, it shows less $Q^2$-dependence
%than it is predicted by perturbative QCD at LO. This slower
%$Q^2$-dependence may be explained naturally by the negative NLO corrections
%to $\sigma$ obtained here.

%\item

\section{Slopes}

The behaviour of eqs. (\ref{9.11}) can mimic a power law shape
over a limited region of $x, Q^2$:
 \begin{eqnarray}
%\z 
f_a(x,Q^2) &\sim &  x^{-\lambda^{eff}_a(x,Q^2)}
~~~~~~ 
~\mbox{ and }~ 
\nonumber \\
F_2(x,Q^2) &\sim &  x^{-\lambda^{eff}_{F2}(x,Q^2)}
\nonumber    
\end{eqnarray}
The quark and gluon effective slopes
 $\lambda^{eff}_a $
%= -\frac{d}{d \ln z} \ln f_a(z,Q^2)$
are reduced by the NLO terms that leads to the decreasing
of the gluon distribution at small $x$. For the quark case
it is not the case, because the normalization factor $A_q^+$ of the ``$+$'' 
component produces an additional contribution undampening as 
$\sim (lnx)^{-1/2}$.

%\vskip -0.5cm

\begin{figure*}[t]
%\rule{5cm}{0.2mm}\hfill\rule{5cm}{0.2mm}
%\vskip -2.5cm
\vskip -0.5cm
%\rule{5cm}{0.2mm}\hfill\rule{5cm}{0.2mm}
%\psfig{figure=filename.ps,height=1.5in}
%\psfig{figure=fi1.ps,height=1.5in}
\psfig{figure=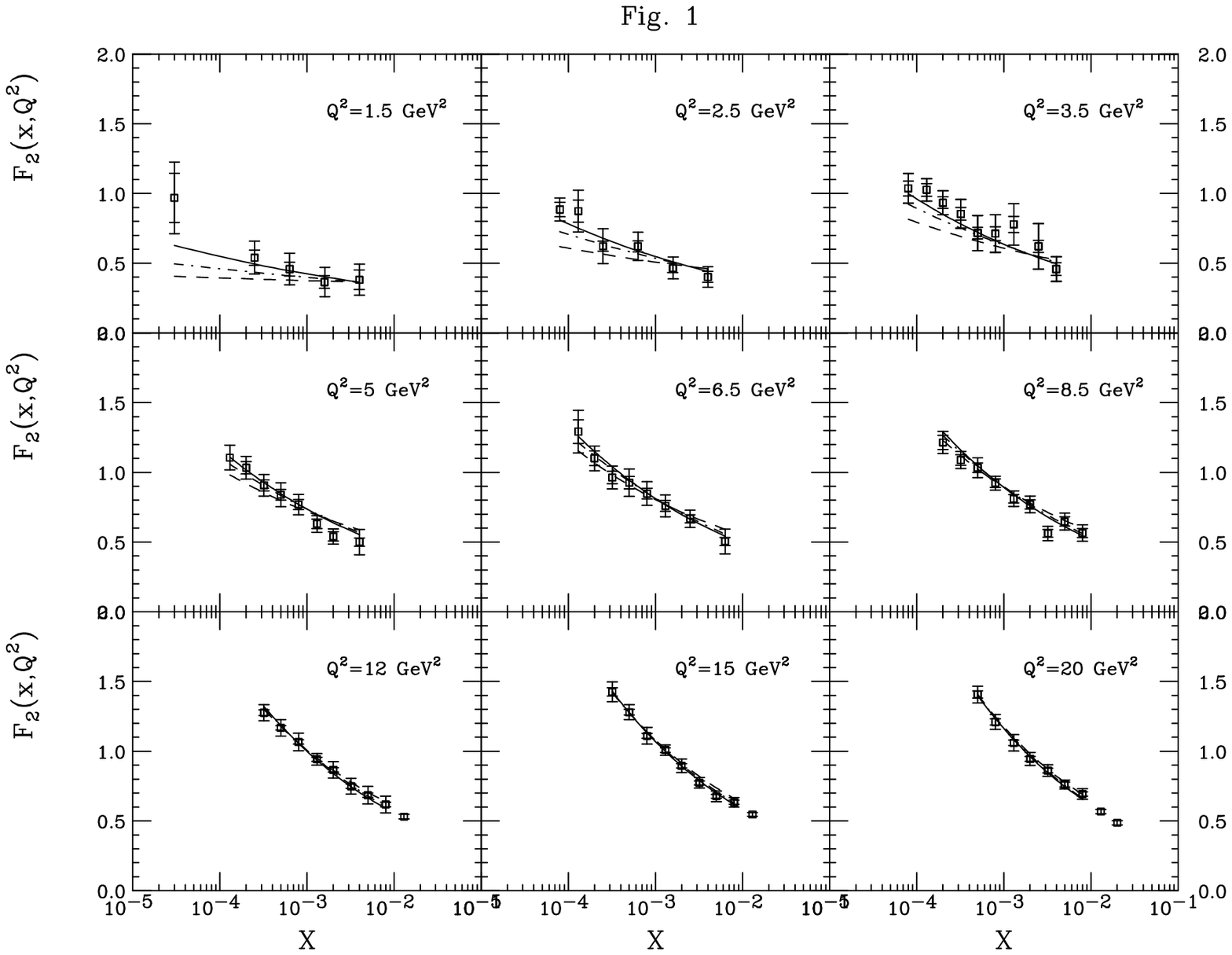,height=2.9in,width=6.2in}
%\psfig{figure=fi1h.ps,height=5.0in,width=5.in}
%\vskip -0.3cm
%\caption{The structure function $F_2$ as a function of $x$ for different
%$Q^2$ bins. The experimental points are from H1. 
%The inner error 
%bars are statistic while the outer bars represent statistic and systimatic 
%errors added in quadrature. The dashen and dot-dashed curves are obtained 
%from fits at LO and NLO respectively with fixed $Q^2_0=1$ GeV$^2$. The solid
%line is from the fit at NLO giving $Q^2_0=0.55$ GeV$^2$.
% $\Lambda_{\ms}(f=4) = 250$ MeV is fixed.}
\vskip -0.5cm
% \label{fig:radish}
\end{figure*}

%\item
The gluon effective slope $\lambda^{eff}_g$ is larger than the quark slope
$\lambda^{eff}_q$, that is in excellent agreement with a recent MRS and 
GRV analyses \cite{MRS}. \\
Indeed,
%because $d/dlnx = d/dlnz$,
the effective slopes
have the
%form,
asymptotical values (at large $Q^2$):
 \begin{eqnarray}
\lambda^{eff,as}_g(x,Q^2) 
%&=& \rho \frac{I_1(\sigma)}{I_0(\sigma)} 
&\approx & \rho - 
\frac{1}{4\ln{(1/x)}} 
\nonumber \\
\lambda^{eff,as}_q(x,Q^2) 
%&=&  \rho \cdot \frac{ I_2(\sigma) (1- \overline d^q_{+-}(1) \alpha(Q^2))
% + 20 \alpha(Q^2) I_1(\sigma)/\rho}{ I_1(\sigma) 
%(1- \overline d^q_{+-}(1) \alpha(Q^2))
% + 20 \alpha(Q^2) I_0(\sigma)/\rho}
% \nonumber \\
&\approx & 
%\biggl( 
\rho - \frac{3}{4\ln{(1/x)}} 
%\biggr) 
%\nonumber \\
%& \cdot &\biggl(1- 
%\frac{10\alpha(Q^2)}{(\hat d_+ s + \hat D_+ p)} \biggr)
%
\label{11.1} \\
\lambda^{eff,as}_{F2}(x,Q^2) 
%
%
%
%
%&=& \lambda^{eff,as}_q(x,Q^2) 
%\frac{ 1 + 6 \alpha(Q^2)/\lambda^{eff,as}_q(z,Q^2)}{ 1 + 
%6 \alpha(Q^2)/\lambda^{eff,as}_g(z,Q^2)} + ~O(\alpha^2(Q^2)) 
%\nonumber \\
&\approx & 
\lambda^{eff,as}_q(z,Q^2) + \frac{3 \alpha(Q^2)}{\ln(1/x)},
\nonumber
\end{eqnarray}
where
symbol $\approx $ marks approximations obtained by expansions of modified
Bessel functions $I_n(\sigma)$. The slope 
$\lambda^{eff,as}_{F2}(x,Q^2)$ lies between quark and gluon ones but
closely to quark slope $\lambda^{eff,as}_{q}(x,Q^2)$ (see also Fig. 2).

%\item
Both slopes $\lambda^{eff}_a$ decrease with decreasing $x$. 
A $x$ dependence of the slope should not appear
for a parton density
%PD 
with a Regge type
asymptotic ($x^{-\lambda}$) and precise measurement of the slope 
$\lambda^{eff}_a$ may lead to the possibility to verify the type of small
$x$ asymptotics of parton distributions.
%
%
%
%\end{itemize}

%%\subsection{Fonts}

\section{Results of the fits}

%\smallskip
%%%  FIGURE  ===  %%%%%%%%%%%%%%%%
   \begin{figure*}[t]
%\label{fig-2}
\unitlength=1mm
\vskip -0.7cm
\begin{picture}(0,70)
  \put(-5,-5){%
   \epsfig{file=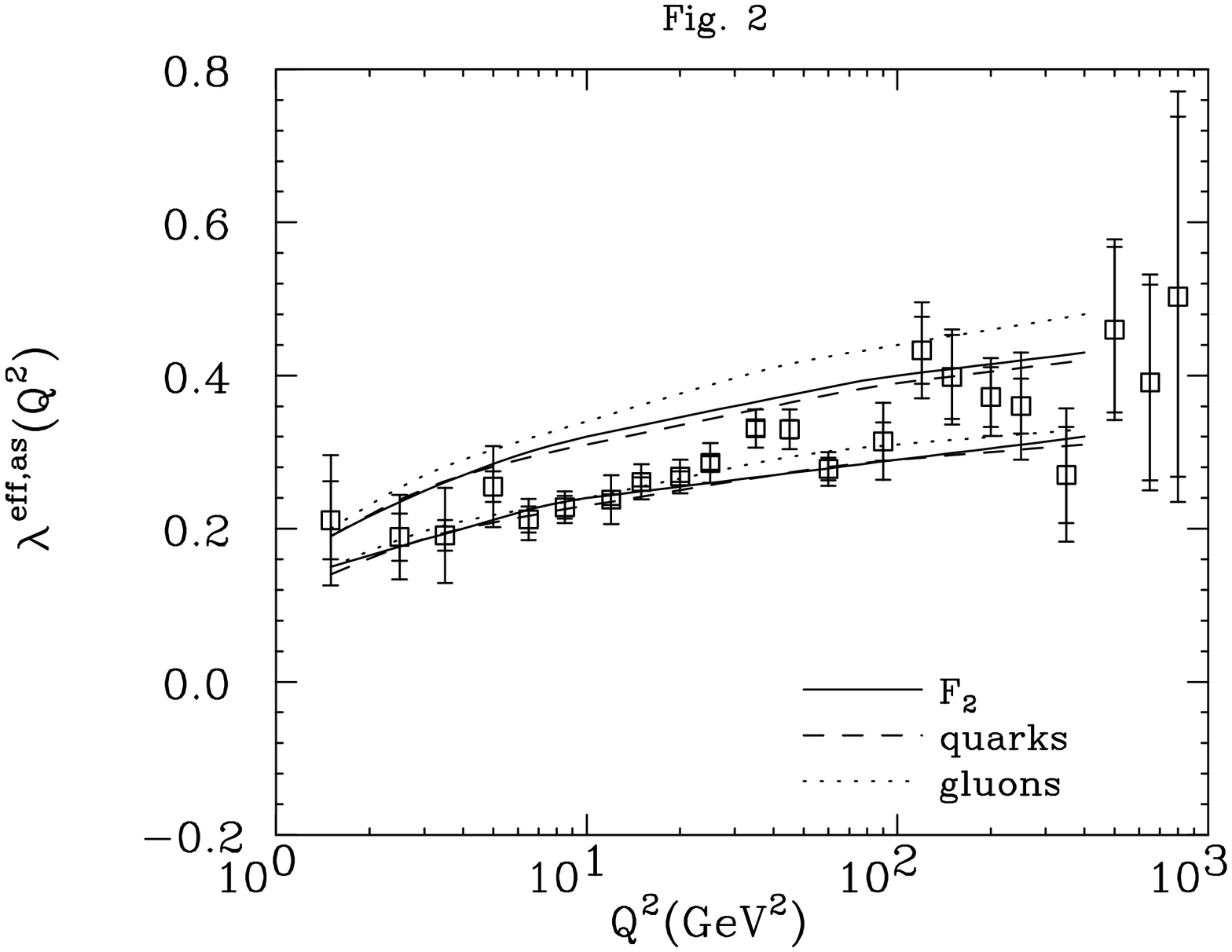,width=80mm,height=180mm,angle=90}%
}
\end{picture}
\vskip -0.3cm
% \caption{\sl  
%Two-loop selfenergy diagrams evaluated in this work.
%Solid lines denote propagators with the mass $M$; dashed lines
%denote massless propagators.}
%\vskip 1cm
 \end{figure*}

With the help of the results obtained in the previous section we have
analyzed $F_2$ HERA data at small $x$ from the H1 collaboration
(first article in \cite{H1}).
%and ZEUS \cite{ZEUS} collaborations separately.
%Initially our solution of the DGLAP equations depends on five
%parameters, i.e. $Q_0^2$, $x_0$, $A_q$, $A_g$ and $\Lambda_{\ms}(n_f=4)$.
In order to keep the analysis as simple as possible
we have fixed $\Lambda_{\ms}(n_f=4) = 250$ MeV which
is a reasonable value extracted from the traditional (higher $x$)
experiments.
%and that has also been used by others \cite{Munich1}.
The initial scale of the 
%PD 
parton densities was also fixed
into the fits to $Q^2_0$ = 1 $GeV^2$, although later it was released
to study the sensitivity of the fit to the variation of this parameter.
The analyzed data region was restricted to $x<0.01$ to remain within the
kinematical range where our results are
accurate. Finally, the number of active flavors was fixed to $f$=4. 

%\verb|\includegraphics{fi1h.ps }|

Fig. 1 shows $F_2$ calculated from the fit
with Q$^2$ $>$ 1 GeV$^2$
%given in table 1
in comparison with H1 data.
Only the lower $Q^2$ bins are shown.
One can observe that the NLO result (dot-dashed line)
lies closer to the data
than the LO curve (dashed line).
The lack of agreement between data and lines observed
at the lowest $x$ and $Q^2$ bins suggests
that the flat behavior should occur at $Q^2$ lower
than 1 GeV$^2$.
In order to study this point we have done the
analysis considering $Q_0^2$ as a free parameter.
Comparing the results of the fits (see \cite{Q2evo})
%in table 3 with those in table 2
one can notice
%a significant reduction in the value of
%$A_g$, $Q_0^2$ and the $\chi^2$. In Fig. 1
the better agreement with the experiment of the
NLO curve at fitted $Q^2_0=0.55 GeV^2$ (solid curve)
is apparent at the lowest kinematical bins.

Finally with the help of Eqs. (\ref{11.1}) we have estimated
the $F_2$ effective slope using the value of the parameters extracted
from NLO fits to data. For H1 data we found
$0.05 < \lambda^{eff}_{F2} < 0.30-0.37$.
%and for ZEUS   
%$0.07-0.09 < \lambda^{eff}_{F2} < 0.31-0.34$.
The lower (upper) limit
corresponds \footnote{For small $Q^2$ we used the exact values of the
slopes presented in \cite{Q2evo}.}
to $Q^2=1.5$ GeV$^2$ ($Q^2=400$ GeV$^2$). The dispersion
in some of the limits is due to the $x$ dependence.
Fig. 2 shows that the three types of asymptotical slopes have
similar values,
which are in very good agreement with H1 data (presented also in Fig. 2).
The NLO values of
$\lambda^{eff,as}_{F2}$ lie between the quark and the gluon ones but
closer to the quark slope $\lambda^{eff,as}_{q}$.
These results are in excellent agreement with those obtained by others 
(see %references \cite{H1,MRS,GRV,Navelet} and also
the review
\cite{CoDeRo} and references therein).

\section{Conclusions} 

We have shown that the results developed recently in \cite{Q2evo}
%As we have shown, these results
 have quite simple form and reproduce many
properties of parton distributions at small $x$,
that have been known from global fits.

We found very good agreement between our approach based on QCD at
NLO approximation and HERA data, as it has been observed earlier with
other approaches (see the review \cite{CoDeRo}). Thus, the nonperturbative
contributions as shadowing effects,
%\cite{Levin},
higher twist effects
%\cite{Bartels}
and others seems to be quite small or seems to be canceled
between them and/or with $ln(1/x)$ terms containing by higher orders of
perturbative theory (see discussion also in \cite{Bartels}).
In our opinion, this very good agreement between 
%our 
approaches based on perturbative QCD 
%at NLO approximation 
and HERA data may be explained also by the fact that
at low $x$ values
the real effective scale of coupling constant
is like $Q^2/x^c$, where $1/2 \leq c \leq 1$
(see \cite{scale}). To clear up the correct contributions of nonperturbative
dynamics and higher orders containing large $ln(1/x)$ terms, it is
necessary
more precise data and further efforts in developing of theoretical
approaches.

  {\it Acknowledgments.}
One of the authors (A.V. K.)
%(A. V. K.)
 would like to express his sincerely thanks to the Organizing
 Committee and especially to R. Fiore and A. Pappa for the kind invitation 
and the financial support
at  such remarkable Conference, and V.S. Fadin, L.L. Jenkovszky
and L.N Lipatov 
%and E.?. Martynov
for fruitful discussions.
%\\
%One of the authors (A.V.K.) 
A.V.K. and G.P. 
were supported in part, respectively, by Alexander von Humboldt fellowship and
RFBR (98-02-16923) and
%and secon one (G.P.) was supported in part 
by Xunta de Galicia
(PXI20615PR) and CICYT (AEN99-0589-C02-02).

\end{document}